\documentclass{aa}
\usepackage{graphicx}
\begin{document}


\title{The $\delta$ Scuti star FG Vir.
V. The 2002 photometric multisite campaign}
 
\author{M.~Breger\inst{1} \and F.~Rodler\inst{1} \and M.~L.~Pretorius\inst{2}
\and S.~Mart\'{\i}n-Ruiz\inst{3}
\and P.~J.~Amado\inst{4}
\and V.~Costa\inst{4}
\and R.~Garrido\inst{4}
\and P.~L\'opez~de~Coca\inst{4}
\and I.~Olivares\inst{4}
\and E.~Rodr\'{\i}guez\inst{4}
\and A.~Rolland\inst{4}
\and T.~Tshenye\inst{5}
\and G.~Handler\inst{1, 6} \and E.~Poretti\inst{3}
\and J.~P.~Sareyan\inst{7} \and M.~Alvarez\inst{8}
\and P.~M.~Kilmartin\inst{9}
\and W.~Zima\inst{1}}

\institute{Institut f\"ur Astronomie der Universit\"at Wien, T\"urkenschanzstr. 17,
A--1180 Wien, Austria\\INTERNET: breger@astro.univie.ac.at
\and
Department of Astronomy, University of Cape Town, Rondebosch 7700, South Africa
\and
INAF -- Osservatorio Astronomico di Brera, Via E. Bianchi 46, I-23807 Merate, Italy
\and
Instituto de Astrof\'{\i}sica de Andaluc\'{\i}a, CSIC, Apdo. 3004, E-18080 Granada, Spain
\and
University of the North West, Mmabatho, South Africa
\and
SAAO, P. O. Box 9, Observatory 7935, Cape Town, South Africa
\and 
Observatoire de la Cote d'Azur, BP 4229, 0634 Nice Cedex 4, France 
\and
Inst. Astronomia IA-UNAM, Apartado 70-264, 04510 Mexico D.F., Mexico
\and
Mount John University Observatory, Department of Physics \& Astronomy,
University of Canterbury, Christchurch, New Zealand}

\date{Received date; accepted date}

\abstract{A high-accuracy multisite campaign was carried out from 2002 January to May
with a photometric coverage of 398 hours at five observatories. The concentration on a few selected sites gives
better consistency and accuracy than collecting smaller amounts from a larger number of sites.
23 frequencies were detected with a high statistical significance.
6 of these are new. The 17 frequencies found in common with the 1992--1995 data
are the modes with highest amplitudes. This indicates that the pulsation spectrum of FG~Vir is relatively
stable over the ten-year period. Two frequencies have variable amplitudes and phases from year to year as well
as during 2002. These were both found to be double modes with close
frequencies. For the mode at 12.15 c/d this leads to an apparent modulation with a time scale of
$\sim$129d. The close frequencies at 12.15 c/d are composed of a radial and a nonradial mode,
suggesting a similarity with the Blazhko Effect seen in RR Lyrae stars.
}
\maketitle

\section{Introduction}

Asteroseismology of $\delta$~Scuti stars has reached a stage where
the choice between different models of stellar structure and
evolution requires a large number of known pulsation frequencies. While
the discovery of many new frequencies cannot be accompanied by successful pulsation mode identifications,
the latter is important to refine the stellar pulsation models. So far, the observations
could not yet be matched with perfect models due, in part, to lack of enough observational
constraints.

Consequently, the Delta Scuti Network (DSN) specializes in the intensive study of a few selected
pulsating stars. One of these objects is the star FG~Vir, which was observed for several months
during 2002. The motivation for additional photometry was:

(i) Successful mode identification relies on both photometric and spectroscopic techniques. While the
first method is used to determine the pulsational $\ell$ values through phase differences and
to a lesser extent, amplitude ratios, the
line-profile technique allows accurate determinations of the $m$ values and determines the rotational
velocity as well as the aspect of the
rotation axis, $i$. In 2002, for the first time the photometric campaign has been paired with
an intensive multisite spectroscopic campaign.

(ii) A comparison of the photometric amplitudes with those obtained during previous campaigns from 1992--1995
would yield valuable information on the long-term stability of the pulsation spectrum.

(iii) Furthermore, a long campaign would lead to a high frequency resolution within the year of observation. This
would permit the search for close frequency pairs, such as those that have been reported for several other $\delta$~Scuti
variables.

(iv) The mode selection mechanism operating in $\delta$~Scuti stars is still unknown.
Theoretical pulsation models predict considerably more modes than are observed. Consequently,
it is important to study the stars in more detail to search for modes with small amplitudes
and to increase the number of known frequencies.

The star FG~Vir (=~HD~106384) has been observed before: its variability was originally
discovered by Eggen (1971) in one night of observation. During 1992,
Mantegazza et al. (1994) measured FG~Vir photometrically for 8 nights.
They were able to identify six frequencies of pulsation,
while a seventh mode of pulsation was also suggested. Two multisite photometric campaigns were
organized by the Delta Scuti Network during 1993 (170 hours, Breger et al. 1995) and 1995 (435 hours,
Breger et al. 1998). The multisite campaigns led to the discovery of 24 frequencies of pulsation, of which
21 were independent pulsation modes. A reasonable agreement between different attempts towards mode identifications
was found (Guzik et al. 1998, Viskum et al. 1998, Breger et al. 1999, Mantegazza \& Poretti 2002).

The Delta Scuti Network is engaged in a long-term study of FG~Vir, using both photometric and spectroscopic techniques.
Consequently, even larger and better analyses are expected in the future. These will include
photometric frequency determinations, photometric and spectroscopic mode identifications as well as stellar modelling.
The purpose of the present paper is to report the 2002 photometry and its implications for our understanding of FG~Vir.

\section{New measurements}

A multisite photometric campaign of FG~Vir was carried out from 2002 January to 2002 May.
DSN campaigns strive for very long observing runs on relatively few telescopes
located on different continents. This maximizes the observational
stability and frequency resolution, while minimizing the negative effects of daily aliasing.
The concentration on a few selected sites gives better consistency and accuracy than collecting
smaller amounts from a larger number of sites. An analyses on the observational accuracy of
global campaigns can be found in Breger (2002).

We consider the evolution of DSN campaigns to fewer sites and longer observations at each site to be
an important step to produce homogeneous, high-quality data.
Furthermore, data from all nights and telescopes with a photometric quality of less than
4 mmag per single observation (as judged from the comparison stars) are rejected.
The journal of usable observations is given in Table~1.

\begin{table*}
\caption[]{Journal of the PMT observations of FG Vir}
\begin{flushleft}
\begin{tabular}{llllllll}
\hline
\noalign{\smallskip}
Start & Length & Observatory & Telescope & \hspace{10mm}Start & Length & Observatory & Telescope\\
HJD (days) & (hours) & & & \hspace{10mm}HJD (days) & (hours) \\
\noalign{\smallskip}
\hline
\noalign{\smallskip}
\noalign{\smallskip}
2452306.60	&	3.6	&	OSN	&	0.9m	&	\hspace{10mm}2452355.45	&	6.0	&	 OSN 	&	 0.9m\\
2452307.64	&	2.9	&	OSN	&	0.9m	&	\hspace{10mm}2452356.46	&	5.6	&	 OSN 	&	 0.9m\\
2452308.60	&	3.0	&	OSN	&	0.9m	&	\hspace{10mm}2452356.71	&	1.8	&	 APT 	&	 0.75m\\
2452310.67	&	1.1	&	OSN	&	0.9m	&	\hspace{10mm}2452357.42	&	6.4	&	 OSN 	&	 0.9m\\ 
2452313.82	&	4.5	&	 APT 	&	 0.75m 	&	\hspace{10mm}2452357.85	&	1.8	&	 APT 	&	 0.75m\\ 
2452314.55	&	4.8	&	OSN	&	0.9m	&	\hspace{10mm}2452358.71	&	5.3	&	 APT 	&	 0.75m\\
2452314.83	&	5.2	&	 APT 	&	 0.75m 	&	\hspace{10mm}2452359.90	&	2.9	&	 MJ 	&	 0.6m \\
2452315.82	&	4.5	&	 APT 	&	 0.75m 	&	\hspace{10mm}2452360.34	&	5.2	&	 SAAO 	&	 0.5m \\
2452316.53	&	5.3	&	OSN	&	0.9m	&	\hspace{10mm}2452363.37	&	4.1	&	 SAAO 	&	 0.5m \\
2452317.81	&	5.2	&	 APT 	&	 0.75m 	&	\hspace{10mm}2452363.74	&	4.4	&	 APT 	&	 0.75m \\
2452324.54	&	4.5	&	OSN	&	0.9m	&	\hspace{10mm}2452364.34	&	5.4	&	 SAAO 	&	 0.5m \\
2452325.57	&	4.0	&	OSN	&	0.9m	&	\hspace{10mm}2452365.35	&	4.7	&	 SAAO 	&	 0.5m \\
2452325.79	&	5.6	&	 APT 	&	 0.75m 	&	\hspace{10mm}2452365.68	&	5.7	&	 APT 	&	 0.75m \\
2452326.79	&	4.5	&	 APT 	&	 0.75m 	&	\hspace{10mm}2452367.31	&	5.8	&	 SAAO 	&	 0.75m \\
2452327.39	&	4.4	&	 SAAO 	&	 0.75m 	&	\hspace{10mm}2452367.68	&	5.6	&	 APT 	&	 0.75m \\
2452327.61	&	1.6	&	OSN	&	0.9m	&	\hspace{10mm}2452368.24	&	5.5	&	 SAAO 	&	 0.75m \\
2452328.55	&	4.6	&	OSN	&	0.9m	&	\hspace{10mm}2452368.68	&	3.9	&	 APT 	&	 0.75m \\
2452328.79	&	4.4	&	 APT 	&	 0.75m 	&	\hspace{10mm}2452369.25	&	1.4	&	 SAAO 	&	 0.75m \\
2452329.52	&	5.4	&	OSN	&	0.9m	&	\hspace{10mm}2452370.35	&	4.8	&	 SAAO 	&	 0.75m \\
2452330.54	&	5.1	&	OSN	&	0.9m	&	\hspace{10mm}2452370.73	&	1.7	&	 APT 	&	 0.75m \\
2452335.77	&	5.3	&	 APT 	&	 0.75m 	&	\hspace{10mm}2452371.24	&	6.7	&	 SAAO 	&	 0.75m \\
2452336.40	&	4.7	&	 SAAO 	&	 0.75m 	&	\hspace{10mm}2452372.66	&	3.1	&	 APT 	&	 0.75m \\
2452336.77	&	5.5	&	 APT 	&	 0.75m 	&	\hspace{10mm}2452373.23	&	2.7	&	 SAAO 	&	 0.75m \\
2452337.44	&	4.0	&	 SAAO 	&	 0.75m 	&	\hspace{10mm}2452373.66	&	3.1	&	 APT 	&	 0.75m \\
2452338.77	&	4.0	&	 APT 	&	 0.75m 	&	\hspace{10mm}2452379.72	&	2.3	&	 APT 	&	 0.75m \\
2452341.59	&	2.5	&	OSN	&	0.9m	&	\hspace{10mm}2452380.45	&	3.7	&	 OSN 	&	 0.9m \\
2452342.48	&	3.5	&	 SAAO 	&	 0.5m 	&	\hspace{10mm}2452380.75	&	3.0	&	 APT 	&	 0.75m \\
2452342.74	&	5.7	&	 APT 	&	 0.75m 	&	\hspace{10mm}2452381.34	&	6.0	&	 OSN 	&	 0.9m \\
2452343.59	&	2.7	&	OSN	&	0.9m	&	\hspace{10mm}2452382.34	&	5.3	&	 OSN 	&	 0.9m \\
2452344.74	&	4.6	&	 APT 	&	 0.75m 	&	\hspace{10mm}2452385.34	&	5.5	&	 OSN 	&	 0.9m \\
2452345.74	&	5.4	&	 APT 	&	 0.75m 	&	\hspace{10mm}2452386.34	&	5.3	&	 OSN 	&	 0.9m \\
2452346.36	&	4.7	&	 SAAO 	&	 0.5m 	&	\hspace{10mm}2452387.39	&	4.2	&	 OSN 	&	 0.9m \\
2452346.73	&	5.7	&	 APT 	&	 0.75m 	&	\hspace{10mm}2452392.35	&	5.0	&	 OSN 	&	 0.9m \\
2452347.40	&	5.0	&	 SAAO 	&	 0.5m 	&	\hspace{10mm}2452394.35	&	4.3	&	 OSN 	&	 0.9m \\
2452347.73	&	0.8	&	 APT 	&	 0.75m 	&	\hspace{10mm}2452395.34	&	4.9	&	 OSN 	&	 0.9m \\
2452348.36	&	6.0	&	 SAAO 	&	 0.5m 	&	\hspace{10mm}2452396.36	&	3.5	&	 OSN 	&	 0.9m \\
2452348.73	&	5.6	&	 APT 	&	 0.75m 	&	\hspace{10mm}2452397.65	&	5.6	&	 SPM 	&	 1.5m \\
2452349.34	&	6.3	&	 SAAO 	&	 0.5m 	&	\hspace{10mm}2452398.47	&	2.3	&	OSN	&	0.9m\\
2452349.84	&	1.4	&	 APT 	&	 0.75m 	&	\hspace{10mm}2452398.67	&	4.0	&	 SPM 	&	 1.5m \\
2452350.38	&	5.3	&	 SAAO 	&	 0.5m 	&	\hspace{10mm}2452399.36	&	3.9	&	OSN	&	0.9m\\
2452351.72	&	5.8	&	 APT 	&	 0.75m 	&	\hspace{10mm}2452399.66	&	5.1	&	 SPM 	&	 1.5m \\
2452352.36	&	5.3	&	 SAAO 	&	 0.5m 	&	\hspace{10mm}2452400.67	&	5.1	&	 SPM 	&	 1.5m \\
2452353.42	&	6.4	&	 OSN 	&	 0.9m 	&	\hspace{10mm}2452402.66	&	5.3	&	 SPM 	&	 1.5m \\
2452353.71	&	5.8	&	 APT 	&	 0.75m 	&	\hspace{10mm}2452403.66	&	1.4	&	 SPM 	&	 1.5m \\
2452354.42	&	6.4	&	 OSN 	&	 0.9m 	&	\hspace{10mm}2452404.72	&	2.6	&	 SPM 	&	 1.5m \\
2452354.72	&	3.6	&	 APT 	&	 0.75m 	\\							
\noalign{\smallskip}
\hline
\end{tabular}
\newline
\end{flushleft}
\end{table*}

The observations were obtained with standard photoelectric
photometers, using photomultiplier tubes as detectors. All measurements
were made through the Str\"omgren $v$ and $y$ filters to provide a relatively
large baseline in wavelength. The three-star technique
in which measurements of the variable star are alternated with those of
two comparison stars, was adopted. Since the same photometric channel is used for all three
measurements, the procedure usually produces the required long-term stability of 3 mmag
or better. We used the same comparison stars as during the 1993 and 1995 campaigns
of FG Vir, viz., HD 106952~(F8V) and HD~105912 (F5V). Again, no variability of these comparison stars
was found.

\begin{figure*}
\centering
\includegraphics*[bb=23 146 564 818,width=176mm,clip]{0830fig1.eps}
\caption{Multisite photometry of FG~Vir obtained during the 2002
DSN campaign. $y$ and $v$ are the observed magnitude
differences (variable -- comparison stars) normalized to zero in the narrowband
$uvby$ system. The fit of the 23-frequency solution derived in this paper is
shown as a solid curve. Triangles: APT; filled diamonds: SAAO 0.5 m;
filled circles: SAAO 0.75 m, open squares: OSN, diamonds with +: MJO, stars: SPM}
\end{figure*}

The following telescopes were used:

\begin{enumerate}
\item The APT measurements were obtained with the T6 0.75~m Vienna Automatic Photoelectric Telescope
(APT), situated at Washington Camp in Arizona. The suitability of this telescope for campaigns
requiring both photometric precision at the millimag level as well long-term stability
was already tested and confirmed by Breger \& Hiesberger (1999).

\item The OSN measurements were obtained with the 0.90~m telescope located
at 2900m above sea level in the South-East of Spain at the Observatorio
de Sierra Nevada in Granada, Spain. The
telescope was equipped with the simultaneous four-channel
photometer ($uvby$ Stromgren photoelectric photometer). The observers were:
P.~Amado, V.~Costa, R.~Garrido, P.~Lopez~de~Coca, S.~Martin~Ruiz; I.~Olivares,
E.~Rodriguez, and A.~Rolland.

\item The SAAO $v$ and $y$ measurements were made with the Modular Photometer
attached to the 0.5~m telescope of the SAAO. The observers were
F.~Rodler and T.~Tshenye. These $v$ and $y$ observations are of the highest
quality with agreement between the comparison stars of $\sim$ 2 mmag.
Additional measurements obtained with the 0.75~m telescope at SAAO by M.~L.~Pretorius and F.~Rodler
were slightly less accurate at the 4 mmag level in $y$, while all $v$ measurements had to be rejected.

\item The SPM measurements were carried out with the 1.5~m telescope
at the San Pedro Martir Observatory, Mexico. The 1.5~m telescope was equipped with a
simultaneous $uvby$ Stromgren photometer, which is the twin of the OSN instrument mentioned
above. The observations from S. Pedro were planned at the end of
the APT campaign in order to minimize overlaps and to significantly increase the time baseline of the
whole DSN campaign. The organization of the measurements and the reductions
were made by E.~Poretti, and the observers were J.~Pierre Sareyan and Manuel Alvarez.
The analysis of the two comparison stars
showed residuals of $\pm$ 3.4 millimag per single observations in the $y$ filter,
while $v$ measurements were of poor quality ($\pm$ 8.4 mmag) due to instrumental problems. 
Consequently, only the (high-quality) $y$ measurements were retained.

\item The MJ measurements supplemented the previous data by an additional night of $y$ measurements
obtained with the 0.6~m telescope at Mount John, New Zealand.
The observer was P.~Kilmartin.

\end{enumerate}

\section{Detection of the pulsation frequencies}

\begin{figure}
\centering
\includegraphics[bb=40 115 555 761,width=85mm, clip]{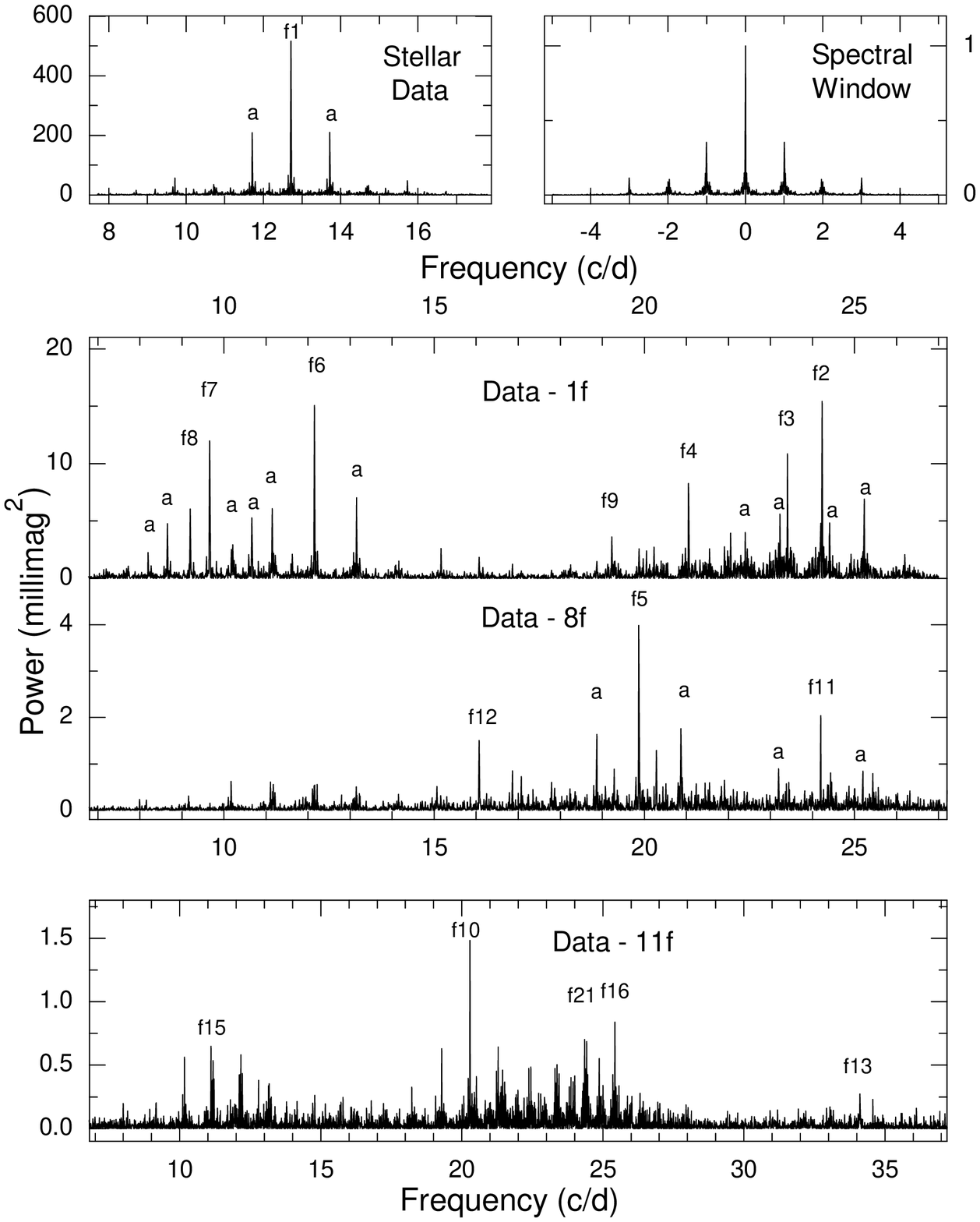}
\caption{Power Spectra of FG Vir for the 2002 photometry.
The numbering scheme (e.g., f$_1$) of the peaks refer to the frequencies found in the
1992--1996 data (Breger et al. 1998). Some 1 c/d alias peaks have been marked by 'a'
for clarity. This figure shows that the 16 pulsation modes with the highest
amplitudes were also present a decade earlier.
Top: The dominant mode and the spectral window. Other panels: Power spectra in the main
pulsation region after prewhitening 1, 8, and 11-frequency solutions.}
\end{figure}

\begin{figure}
\centering
\includegraphics[bb=29 122 542 608,width=85mm, clip]{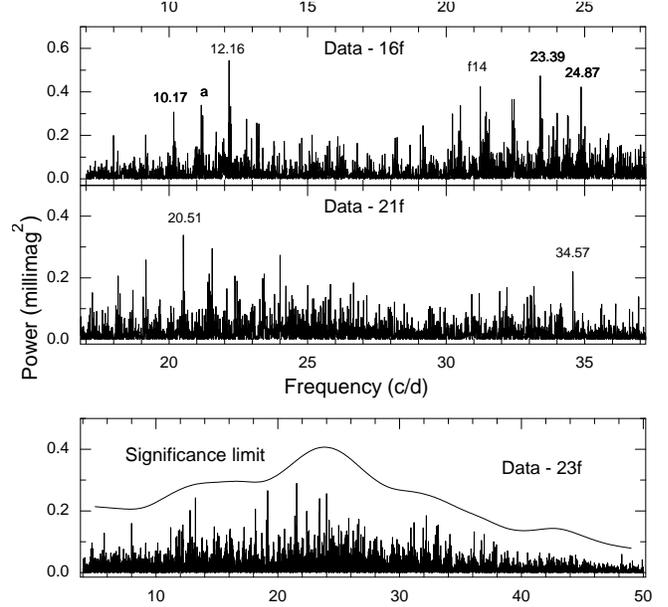}
\caption{Additional pulsation modes detected in the 2002 data after
prewhitening 16 previously known frequencies. The values
(in c/d) of the new modes are shown. Otherwise the symbols have the same meaning as in Fig. 2.}
\end{figure}

The pulsation frequency analyses were performed with a package of computer
programs with single-frequency and multiple-frequency techniques (programs
PERIOD, Breger 1990; PERIOD98, Sperl 1998), which utilize Fourier as well as
multiple-least-squares algorithms. The latter technique fits a number of
simultaneous sinusoidal variations in the magnitude domain and does not rely
on prewhitening.

To decrease the noise in the power spectra, we combined the $y$ 
(4393 new measurements) and $v$ data (4147 new measurements). The dependence of the pulsation amplitude
on wavelength was compensated by multiplying the $v$ data set by an experimentally
determined factor of 0.70 and increasing the weight of these data points by 1/0.70.
This scaling creates similar amplitudes but does not falsify the power spectra.
Note that different colors and data sets were only combined to detect new frequency peaks
in the Fourier power spectrum and to determine the significance of the detection.
The effects of imperfect amplitude scaling and small phase shifts between colors
can be shown to be very small. For prewhitening, separate solutions were obtained for
each color by multiple least-square fits (PERIOD98).

No special weighting of the data points according to their accuracy was applied, except for
assigning a weight of 0.25 to 3 nights of the SAAO 75-cm telescope measurements (dates 245 2327 to
245 2336) and the 5 last OSN nights (JD 245 2387 to 245 2396). The measurements were of slightly
lower precision (see previous section).

In the analysis of the Delta Scuti Network campaign data, we usually apply
a specific statistical criterion for judging the reality of a newly discovered peak in the
Fourier spectra, viz., a ratio of amplitude signal/noise = 4.0 (see Breger et al. 1993).
For FG~Vir, previous campaigns have led to the discovery of 24 significant and
8 probable frequencies. In order to examine the long-term stability of the pulsation of FG~Vir
we carried out an independent analysis of the new data to see which of the previously known
pulsation modes could be detected in the new data. 

Our analysis involves a number of different steps to be repeated. Each step involves the computation
of a Fourier analysis (power spectrum) from the original data or a previously prewhitened fit.
The dominant peaks in the power spectrum were then examined for statistical significance and possible
effects of daily and annual aliasing. For computing new multifrequency solutions, the amplitudes and phases were computed
separately for each color, so that even these small errors associated with combining
different colors were avoided. Note that the new multifrequency solutions always
were computed from the observed (not the
prewhitened) data. Because of the day-time and observing-season (annual) gaps,
different alias possibilities were tried out and the fit with the lowest residuals selected.
The resulting optimum multifrequency solutions were then prewhitened and the analysis repeated
while adding more and more frequencies, until the new peaks were no longer statistically significant.

Figures 2 and 3 show the results of the frequency search. No statistically significant peaks
were found in the low-frequency region, which are therefore not shown in the figures. The question
arises whether the adopted reduction procedures could have cancelled low-frequency terms intrinsic
to the star. The data from the different telescopes were merged by adopting an artifical zero-point for
each telescope: for the initial analysis the average magnitude was computed for each telescope and for
later analyses the zero-point of the multifrequency solution was adjusted. This approach can, in principle,
suppress real low-frequency power, especially near 1.00 c/d. However, our data were gathered essentially
with only four telescopes so that each data set contains enough measurements to detect low frequencies.
Numerical simulations have shown that our reduction procedure should not have negatively influenced the
low-frequency analyses. This potential problem also shows that multisite campaigns with large amounts of data
gathered from few sites are more accurate than a collection from a large number of different telescopes.

Altogether 23 statistically significant peaks in the range of 9.2 to 34.6 c/d were found in the 2002 data.
17 of these were previously known. Of the six new frequencies, one peak is a combination frequency:
the 24.87 c/d peak can be identified with $f_1$+$f_6$, which are the modes with the highest photometric
amplitudes. Two of the new modes are the secondary components of mode doublets (see next section).

The results of the search for multiple frequencies in the 2002 FG~Vir data are shown in Table~2.
Since the third decimal place of the frequencies is uncertain,
only two decimal places are given. Exceptions are the close pairs.

We also list the seven previously announced frequencies not found to be significant in the 2002 data.
Some may be present in the data, but do not reach the signal/noise criterion for a secure detection.
In order to estimate their amplitude, we have reduced the noise by combining the two filters, $v$ and
$y$ after scaling the $v$ data by 0.70 in order compare with $y$.
We have computed their amplitudes in an additional multifrequency solution including these frequencies
and listed in the amplitudes in brackets. We note that $f_{24}$ is not present in the data.

\begin{table}
\caption{Amplitudes of the multiple frequencies of FG Vir}
\begin{tabular}{llccl}
\hline
\noalign{\smallskip}
\multicolumn{2}{c}{Frequency$^{1}$}& \multicolumn{2}{c}{Amplitudes  
(2002)} & Comments\\
& & $v$ & $y$ \\
& c/d & mmag & mmag\\
\noalign{\smallskip}
\hline
\noalign{\smallskip}
$f_1	$ &	12.72	&	32.3	&	22.1	\\
$f_2	$ &	24.23	&	5.7	&	4.2	\\
$f_3	$ &	23.403	&	5.9	&	4.3	& doublet \\
new     & 23.397  &   1.3 & 0.8  & doublet \\
$f_4	$ &	21.05	&	4.5	&	3.1	\\
$f_5	$ &	19.87	&	3.0	&	2.1	\\
$f_6	$ &	12.154	&	6.1	&	4.2	& doublet\\
new     & 12.162  &   1.3 & 1.0 & doublet\\
$f_7	$ &	9.66	&	5.3	&	3.7	\\
$f_8	$ &	9.20	&	3.8	&	2.8	\\
$f_9	$ &	19.23	&	2.5	&	1.6	\\
$f_{10}	$ &	20.29	&	1.8	&	1.1	\\
$f_{11}	$ &	24.19	&	2.3	&	1.7	\\
$f_{12}	$ &	16.07	&	1.6	&	1.3	\\
$f_{13}	$ &	34.12	&	0.7	&	0.6	\\
$f_{14}	$ &	21.23	&	1.0	&	0.6	\\
$f_{15}	$ &	11.10	&	0.8	&	0.7	\\
$f_{16}$ &	25.43	&	1.3	&	0.9 & = 2$f_1$\\
$f_{21}	$ &	24.35	&	1.1	&	0.9	\\
new & 10.17 & 0.8 & 0.5\\
new & 24.87	& 0.8 & 0.4	  & = $f_1$ + $f_6$\\
new & 20.51 & 0.6 & 0.7\\
new & 34.57 & 0.6 & 0.4\\
\noalign{\smallskip}
\multicolumn{2}{l}{Residuals}&$\pm$ 4.1 & $\pm$ 3.7\\
\noalign{\smallskip}
\multicolumn{5}{l}{Previous frequencies not found to be significant$^{2}$ }\\
$f_{17}$ &	33.06 & & (0.3)\\
$f_{18}$ &	21.55 & & (0.4)\\
$f_{19}$ &	28.14 & & (0.4)\\
$f_{20}$ &	11.20 & & (0.4)\\
$f_{22}$ &	11.87 & &(0.3)\\
$f_{23}$ &	22.37 & & (0.4) & = $f_1$ + $f_7$\\
$f_{24}$ &	10.69 & & (0.1) & = $f_1$ - $f_3$\\
\noalign{\smallskip}
\hline
\noalign{\smallskip}
\end{tabular}
\newline
$^{1}$ The 'old' frequency notation from Breger et al. (1998)\\
was used.\\
$^{2}$ Frequencies not found to be significant may be excited\\
with small amplitudes, but the existence of $f_{24}$ is not\\
confirmed. They were not used for the final solution\\
and residuals.\\
\end{table}

The power spectrum of the residuals (after subtracting a 23-frequency fit) contains additional information too.
If all the pulsation modes on the star were found by us, the noise figure should show a steady, slow decrease
with frequency. This is not seen. Inspection of the power spectrum indicates that
many additional pulsation modes {\it{in the same frequency bands as the detected modes}} are present.
These are the 11-13, 19-25 and 30-35 c/d regions. The significance limit adopted by us (4 x average amplitude of
the noise sampled over 4 c/d
regions) measures both observational errors and undetected peaks. Our previous applications of this limit
to multisite data suggests that it is quite conservative and avoids incorrect frequency detections.
The latter is important since stellar pulsation models attempt to fit every detection, so that
incorrect frequencies are more harmful than undetected additional modes.

We conclude that while the present campaign has detected six additional modes,
more modes are present in the star with amplitudes of 0.5 mmag or lower.

\section{Close frequency doublets and amplitude variability}

It was shown by Breger \& Bischof (2002) that the majority of the well-studied
$\delta$~Scuti stars show close frequency pairs. With insufficient
frequency resolution this would resemble amplitude variability of a single mode.
A good example is offered by the 8.65 c/d mode(s) in BI~CMi (same paper), where two long
observing seasons independently established a close frequency pair separated by
0.02 c/d as the appropriate explanation.

For FG~Vir, we already noted the stability of the pulsation modes between 1992--1995
and 2002. A few modes, however, show amplitude changes. Breger et al. (1998) already noted the
strong amplitude variability of the 23.4 c/d peak from 1992 to 1995, although the annual
coverage was too short to look for short-term changes within an observing season.
However, the 98d coverage of the 2002 data allows the detection of amplitude and period (phase)
changes within the single year for the 12.15 c/d and 23.4 c/d peaks.
Two probable explanations are: (i) two modes with close frequencies beating with each other, and (ii) a
single mode with variable frequency and amplitudes. This problem was already examined
in our analysis of BI~CMi (Breger \& Bischof 2002), where for a number
of modes it was possible to choose between the
two interpretations on the basis of a specific test in which the amplitude changes and associated
phase changes of an assumed single mode were examined. This test works well when the amplitudes
of the two close modes are similar. For BI~CMi, the variable-amplitude single-frequency model could
be rejected. For FG~Vir, the phase test is not very helpful because of the very different amplitudes.

Let us examine the mode(s) near 12.15 c/d.
The power spectra of the 2002 data are shown in Fig.~4.
After the main mode
at 12.154 c/d is prewhitened, we are left with a single peak at 12.162 c/d. Prewhitening both
peaks leaves only noise. This provides strong evidence that two separate frequencies
beating with each other are involved and that amplitude variability of a single mode
is not responsible. However, the changes are at the limit or resolution for the 2002
data: the conclusion could be strengthened if a separate observing
season would yield the same two peaks, thereby eliminating the possibility that
the time coverage of the data conspired to suppress the additional
peaks required by pure amplitude variability. We note that the two-frequency solution removes
the observed amplitude and phase variability.

The main mode at 12.15 c/d has been shown to be radial on the basis of phase differences
between the $v$ and $y$ light curves (Breger et al. 1999) and equivalent-width changes (Viskum et al. 1998).
The secondary mode must, therefore, be nonradial because the period ratios of radial modes do
not permit such close spacing. A nonradial mode with a frequency close to the frequency of the radial mode
is one of the promising explanations for the observed amplitude variations
associated with the Blazhko Effect in RR Lyrae
stars (Kolenberg et al. 2003, Kovacs 2002). We speculate that the two phenomena in RR Lyrae and $\delta$~
Scuti stars may have a similar astrophysical origin.

The observational results for the 23.40 c/d modes are similar except that here a beat cycle of $\sim$163d, instead
of the 129d for 12.15 c/d modes, is calculated. Again, this needs to be confirmed in data with even higher frequency resolution.

\begin{figure}
\centering
\includegraphics[bb=44 234 547 744,width=85mm, clip]{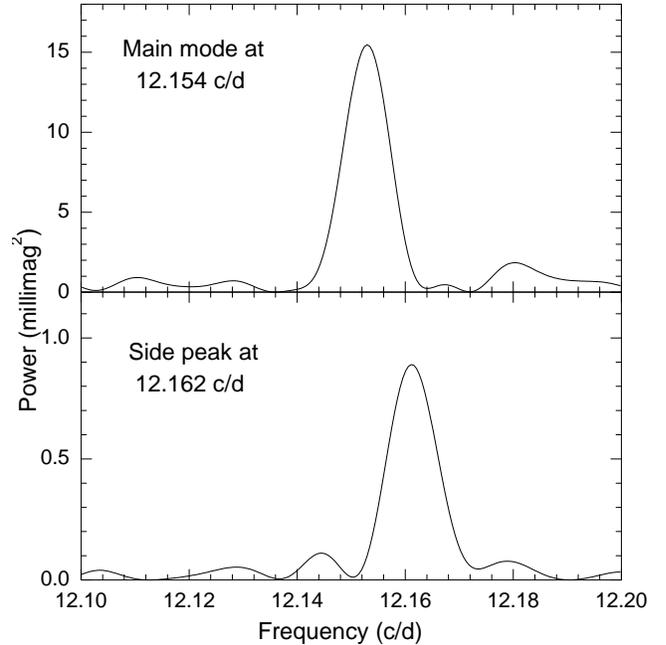}
\caption{Power spectrum of the radial 12.15 c/d mode (see text for
filter information). In the bottom panel
the radial mode has been prewhitened. Note the different scales in power.
This figure shows that the
observed amplitude variability of the radial mode can be caused by
the presence of a close nonradial mode.}
\end{figure}

\section{Discussion}

16 out of the 17 frequencies found to be statistically significant in both the 1992--1995 and 2002 data
are the modes with the largest photometric amplitudes. While this is hardly surprising, it also provides an argument
in favor of the stability of the pulsation spectrum of FG~Vir. In this regard the star differs from some other
$\delta$~Scuti stars such as 4~CVn (see Breger 2000), in which measurements taken ten years apart may even provide
a first impression of belonging to two different stars! Actually, in 4~CVn, the 'disappeared' modes have been
shown to be still present, but with much smaller amplitudes.

The 'missing' modes in FG~Vir all had $V$ amplitudes between 0.4 and 0.8 mmag during 1995 with
a statistical uncertainty of $\pm$ of 0.11 mmag. The 2002 data
show that most may still be present with the peaks at the correct frequencies, but with amplitudes less than 0.5 mmag.
This is below the significance limit. We conclude that the differences can be explained by observational
uncertainties in most cases.

\acknowledgements

This investigation has been supported by the
Austrian Fonds zur F\"{o}rderung der wissenschaftlichen Forschung,
project number P14546-PHY. The Sierra Nevada observations
were supported by the Junta de Andalucia and the DGI under project AYA2000-1580.
PJA acknowledges financial support at the Instituto de
Astrof\'{\i}sica de Andaluc\'{\i}a-CSIC by an I3P contract
(I3P-PC2001-1) funded by the European Social Fund.
SMR also acknowledge the financial support by a Marie Curie
Fellowship of the European Community programme 'Improving the Human
Potential' under contract number HPMF-CT-2001-01146.

\end{document}